\documentclass{aa}
\usepackage[varg]{txfonts}
\usepackage[colorlinks=true,linkcolor=blue,citecolor=blue]{hyperref}
\usepackage{graphicx}
\usepackage{color}

\begin{document} 

 \title{Active minimization of non-common path aberrations in long-exposure imaging of exoplanetary systems}
 \titlerunning{Active minimization of non-common path aberrations}
\authorrunning{Singh et al.}
  \author{Garima Singh \inst{}, Rapha\"el Galicher \inst{}, Pierre Baudoz \inst{}, Olivier Dupuis \inst{}, Manuel Ortiz \inst{}, Axel Potier \inst{}, Simone Thijs \inst{} \and Elsa Huby \inst{}}

  \institute{LESIA, Observatoire de Paris, Universit\'e PSL, CNRS, Universit\'e de Paris, Sorbonne Universit\'e, 5 place Jules Janssen, 92195 Meudon, France\\
             \email{garima.singh@obspm.fr}} 
            

  \abstract
   {Spectroscopy of exoplanets is very challenging because of the high star-planet contrast. A technical difficulty in the design of imaging instruments is the noncommon path aberrations (NCPAs) between the adaptive optics (AO) sensing and the science camera, which induce planet-resembling stellar speckles in the coronagraphic science images. In an observing sequence of several long exposures, quickly evolving NCPAs average out and leave behind an AO halo that adds photon noise to the planet detection. Static NCPA can be calibrated a posteriori using differential imaging techniques. However, NCPAs that evolve during the observing sequence do not average out and cannot be calibrated a posteriori. These quasi-static NCPAs are one of the main limitations of the current direct imaging instruments such as SPHERE, GPI, and SCExAO.}
   {Our aim is to actively minimize the quasi-static speckles induced in long-exposure images. To do so, we need to measure the quasi-static speckle field above the AO halo.}
   {The self-coherent camera (SCC) is a proven technique which measures the speckle complex field in the coronagraphic science images. It is routinely used on the THD2 bench to reach contrast levels of $<10^{-8}$ in the range $5-12\,\lambda/D$ in space-related conditions. To test the SCC in ground conditions on THD2, we optically simulated the residual aberrations measured behind the SPHERE/VLT AO system under good observing conditions.}
   {We demonstrate in the laboratory that the SCC can minimize the quasi-static speckle intensity in the science images down to a limitation set by the AO halo residuals. The SCC reaches 1$\sigma$ raw contrast levels below $10^{-6}$ in the region $5-12\,\lambda/D$ at 783.25~nm in our experiments.}
 {The results presented in this article reveal an opportunity for the current and future high-contrast imaging systems to adapt the SCC for real-time measurement and correction of quasi-static speckles in long-exposure science observations from the ground.}


   \maketitle
\section{Introduction}
\label{s:intro}

One of the goals of the Extremely Large Telescopes (ELTs) is to study the composition and chemistry of the exoplanetary atmospheres using direct imaging. Such a challenge requires the star-planet light to be separated. Exoplanets are $10^{4}$ to $10^{10}$ times fainter than their host stars at only a fraction of an arcsecond, and a device called a coronagraph is required to reduce the stellar flux without affecting the light from the planet. The efficiency with which a coronagraph can reject the starlight depends on how well a high-contrast imaging (HCI) instrument deals with the optical aberrations of various natures \citep[atmospheric turbulence and optical defects;][]{guyon2005, guyon2006}. The atmosphere of Earth introduces dynamic wavefront errors creating quickly varying speckles in coronagraphic science images obtained by the ground-based HCI instruments. The adaptive optics (AO) systems compensate most of these fast wavefront errors, but the AO correction is not perfect. Servo lag errors and noncommon path aberrations (NCPAs) between the sensing and the science channel degrade the performance of AO systems \citep{sdi, bloe, roddier_book}. Despite using the best AO systems \citep{gems2016, fusco2016}, speckle patterns appear in coronagraphic images and readily imitate signals from exoplanets \citep{beuzit1997, oppen2001, bocc2003, bocc}. 

In a typical long exposure, the AO-induced quickly varying speckles (due to servo lag errors) average out and create a low spatial frequency halo that adds photon noise when looking for point-like sources such as exoplanets. Static speckles in an observing sequence of such exposures can be calibrated post-observation using differential imaging techniques \citep{sdi, marois_adi, mar2, mar1}. Quasi-static speckles however evolve from one exposure to  another with the slowly changing aberrations (thermal changes and gravity flexures). They can partially be calibrated a posteriori by differential imaging techniques, however the residuals left behind limit all the current HCI instruments such as SPHERE \citep[Beuzit et al. 2019 under review]{Beuzit06}, GPI \citep{gpi_firstlight}, and SCExAO \citep{nem_scexao}. In conclusion, the long-exposure coronagraphic images coming out of these instruments have three main visible features: a smooth halo, static speckles that remain unchanged in all images of a science sequence, and quasi-static speckles that vary slowly during the sequence. While we consider in our paper that the lifetime of static and quasi-static speckles is longer than the exposure time (typically a few seconds), it is assumed that the evolving time of quasi-static speckles is shorter than the length of the observing sequence (typically one hour). Our focus is therefore to address the quasi-static speckles that remain static for a few images but vary along the complete sequence of observations. 

With a priori knowledge of speckle evolution lifetime \citep{mili}, more evolved a posteriori algorithms may well calibrate the speckle pattern. However, any such method can directly benefit from an active technique that minimizes the static or quasi-static speckles in each science image during an observation. Active suppression of these speckles requires measurement of the electric field associated with the speckles directly from a coronagraphic image using a focal plane wavefront sensor (FPWFS). Several FPWFSs have been proposed such as phase diversity \citep{borde2006, efc, sauvage2012} and the self-coherent camera \citep[SCC,][]{scc1,sc}. Once the electric field is measured, one or several deformable mirrors (DM) can then be used to minimize the speckle intensity in a region of the image called the dark hole \citep{malbet95}. Very encouraging laboratory results have been obtained on the coronagraphic testbeds simulating the space-related environment. The stellar speckle intensity is shown to be reduced by a factor of up to~$10^5$ \citep{belikov2007, trauger2011, trauger2012, scc3, scc4} which would enable the detection of planets $10^{10}$ times fainter than their host star. Several attempts on ground-based Extreme-AO instruments have been performed with moderate results. The stellar speckle intensity has been suppressed by only a factor of up to~ten \citep[Galicher et al. in Prep]{sav2012, sn, bot, matthews2017, wilby2017, vigan2019} reaching contrast levels of roughly $10^{-6}$. Most of these techniques temporally modulate the speckle intensity to measure their phase and at least three images are needed for each estimation of the electric field. The quasi-static speckles that evolve faster than every four images cannot be correctly estimated and therefore set a limitation for the speckle estimation. Addressing this concern, our team proposed the SCC that spatially modulates the speckle intensity so that the associated complex electric field can be measured in every science image. The drawback is that a finer sampling of the coronagraphic image is required as compared to the temporal modulation techniques. The SCC has been developed and rigorously tested in space-related environments on the THD2 bench at the Paris Observatory \citep{baudoz2018}.

This paper presents the laboratory performance of the SCC in ground-based conditions. In section~\ref{s:wave}, we first define the expression of the averaged electric field that can be measured in a long-exposure coronagraphic science image given any FPWFS. Section \ref{s:resultNum} presents numerical simulations showing the expected level of quasi-static aberrations that can be measured on THD2 for a given exposure time. In section~\ref{s:thd2}, the components of the THD2 bench and the installation of a subsystem that mimics SPHERE AO residuals under good observing conditions are described. Section \ref{s:scc} reiterates the concept of SCC and describes how the sensor estimates the complex electric field from a long-exposure coronagraphic image. Section \ref{s:resultExp} presents the closed-loop laboratory performance and contrast results. 

\section{Expression of the complex electric field in long exposure}
\label{s:wave} 
\begin{figure}
   \includegraphics[width=9cm]{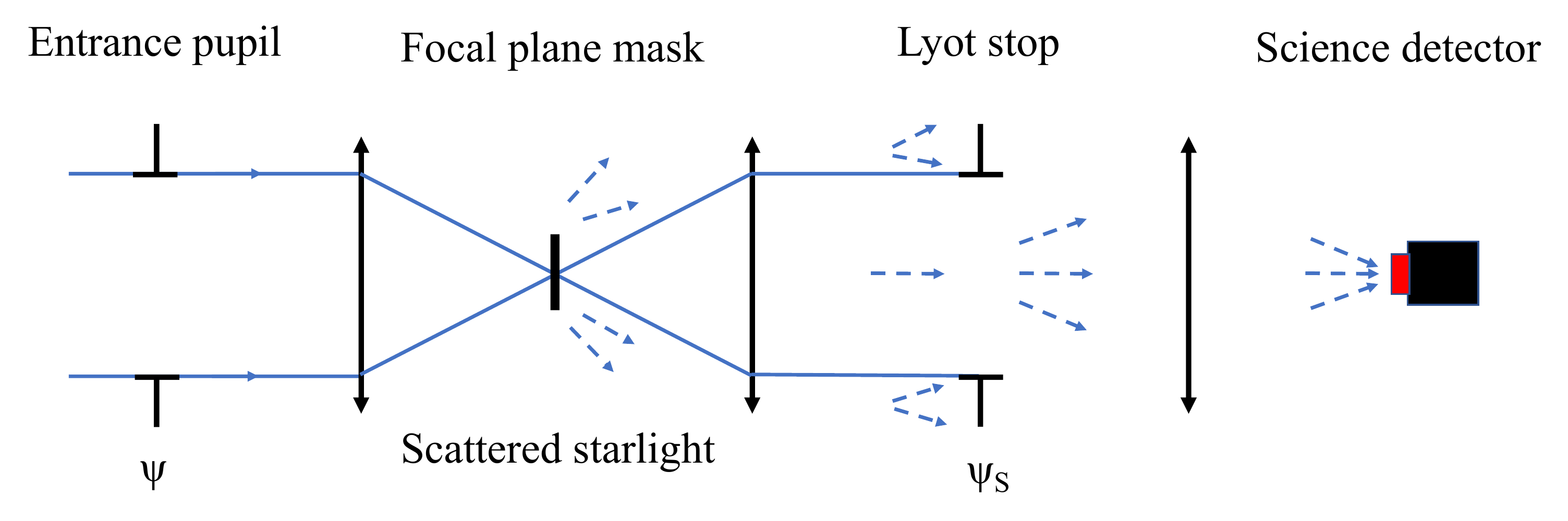}
      \caption{Optical layout of a typical coronagraphic system.}
         \label{fig:coro_layout_1}
   \end{figure}

In this section, we establish an expression of the electric field in long-exposure images of a typical coronagraphic system as shown in Fig.~\ref{fig:coro_layout_1}. The star is assumed to be a monochromatic unresolved source centered on the optical axis. The starlight goes through the entrance pupil and is diffracted by a coronagraphic mask at the focal plane. This scattered starlight downstream of the mask is blocked by a Lyot stop. In the case of aberrations, part of the scattered starlight leaks through the Lyot stop and creates speckles at the final focal plane where a science detector is placed. For simplicity, we do not focus on deriving a classical expression of the complex electric field at each step of a coronagraphic system. Instead, we are interested in the electric field at the Lyot plane that can be linked to aberrations in the entrance pupil.

During an exposure time~$t$, the electric field $\Psi(\vec{\boldsymbol{\xi}},\,t)$ at the entrance pupil in Fig.~\ref{fig:coro_layout_1} in the presence of static aberrations~$\phi_0$ and independently evolving aberrations~$\phi_1(t)$ with a lifetime of~$t_1$ can be represented as
\begin{equation}
\Psi(\vec{\boldsymbol{\xi}},\,t) = P_0(\vec{\boldsymbol{\xi}})\,e^{\displaystyle i\,\left[ \phi_0(\vec{\boldsymbol{\xi}}) +\phi_1(\vec{\boldsymbol{\xi}},\,t)\right]}  ,
\label{eq:psi_entrance}
\end{equation}
where~$P_0$ describes the entrance geometrical pupil and $\vec{\boldsymbol{\xi}}$ is the position in the pupil plane. Here, a perfect coronagraph is considered, which completely rejects the starlight and contributes to the null electric field inside the Lyot pupil in the absence of aberrations. In the case of static aberrations ($\phi_0$) only, the field~$\Psi_S(\vec{\boldsymbol{\xi}})$ at a position~$\vec{\boldsymbol{\xi}}$ in the pupil plane downstream of a perfect coronagraph \citep[and Appendix~\ref{eq:A3}]{cavarroc06} can be represented as
\begin{equation}
\Psi_S(\vec{\boldsymbol{\xi}}) = P_0(\vec{\boldsymbol{\xi}})\,\left(e^{\displaystyle i\,\phi_0(\vec{\boldsymbol{\xi}})}-e^{\displaystyle-\sigma_0^2/2}\right), 
\label{eq:psi_s_stat}
\end{equation}
 where $\sigma_0^2$ is the spatial variance of~$\phi_0$. It has been demonstrated in numerical simulations and in the laboratory that this equation can be used to estimate and control the static aberrations~$\phi_0$ in short exposures~\citep{scc2,scc3,scc4}.

Our objective is to measure~$\phi_0$ from a long-exposure. We assume that during this exposure, a part of the aberration called $\phi_1(t)$  evolves quickly, with a lifetime~$t_1$ (post-AO residuals for example). In such a case, the instantaneous field~$\Psi_S$ in the pupil plane after a perfect coronagraph at a time~$t$ can then be written as
\begin{equation}
\Psi_S(\vec{\boldsymbol{\xi}},\,t) = P_0(\vec{\boldsymbol{\xi}})\,\left(e^{\displaystyle i\,\left[ \phi_0(\vec{\boldsymbol{\xi}}) +\phi_1(\vec{\boldsymbol{\xi}},\,t)\right]}-e^{\displaystyle-(\sigma_0^2+\sigma_1^2)/2}\right),
\label{eq:psi_s}
\end{equation}
where~$\sigma_1^2$ is the spatial variance of~$\phi_1(\vec{\boldsymbol{\xi}},\,t)$, which is considered constant over time. For an infinite exposure time, the electric field is the mathematical expectation of~$\Psi_S$ over time (see Appendix~\ref{eq:A4}). It is represented as
\begin{equation}
E\left[\Psi_S(\vec{\boldsymbol{\xi}},\,t)\right] = P_0(\vec{\boldsymbol{\xi}})\,\left(e^{\displaystyle i\,\phi_0(\vec{\boldsymbol{\xi}})}-e^{\displaystyle -\sigma_0^2/2}\right)\,e^{\displaystyle -\sigma_1^2/2}.
\label{eq:psi_s_inf}
\end{equation}
Here, it is assumed that the variance of~$\phi_1$ over time is equal to the variance over space. The expression in Eq.~\ref{eq:psi_s_inf} is the same as in the case of static errors only (Eq.~\ref{eq:psi_s_stat}) multiplied by a constant factor~$e^{\sigma_1^2/2}$. The Eq.~\ref{eq:psi_s_inf} interprets that the evolving aberration~$\phi_1$ averages out and the static aberrations~$\phi_0$ can be measured from an infinite exposure.

For a finite exposure which is~$\tilde{N}$ times longer than the aberration lifetime~$t_1$, the field~$<\Psi_S(\vec{\boldsymbol{\xi}},\,t)>_{\tilde{N}}$ is the average over~$\tilde{N}$ independent~$\phi_1$ aberrations and can be written as 
\begin{equation*}
<\Psi_S(\vec{\boldsymbol{\xi}},\,t)>_{\tilde{N}} = \frac{1}{\tilde{N}}\,\sum_{p=1}^{p=\tilde{N}}\Psi_S(\vec{\boldsymbol{\xi}},\,p\,t_1).
\end{equation*}
If~$\tilde{N}$ tends to infinity, the above equation is equal to~$E\left[\Psi_S(\vec{\boldsymbol{\xi}},\,t)\right]$. If~$\tilde{N}$ is a finite number, the part of the field that is not averaged is given by the variance of~$<\Psi_S(\vec{\boldsymbol{\xi}},\,t)>_{\tilde{N}}$ over time (see Appendix~\ref{eq:A8}):
\begin{equation}
    \mathrm{Var}\left[<\Psi_S(\vec{\boldsymbol{\xi}},\,t)>_{\tilde{N}}\right]= P_0(\vec{\boldsymbol{\xi}})\,\frac{1-e^{\displaystyle -\sigma_1^2}}{\tilde{N}}.
    \label{eq:psi_s_fin_var}
\end{equation}
We note here that the lifetime~$t_1$ of~$\phi_1$ can be different for each spatial frequency. A comprehensive formalism that includes this assertion will be presented in a future publication. Here, we focus on the aberration~$\phi_1$ which is dominated by two spatial frequencies:~$20$\,cycles per pupil diameter (AO cut-off) and less than~$2$ cycles per pupil diameter (see section~\ref{s:resultNum} and appendix).

We consider small aberrations in the experiments presented in this paper and~rewrite Eqs.~\ref{eq:psi_s_inf} and~\ref{eq:psi_s_fin_var} as follows:
\begin{equation*}
\left \{
\begin{array}{c @{\quad\simeq\quad} c}
    E\left[<\Psi_S(\vec{\boldsymbol{\xi}},\,t)>_{\tilde{N}}\right] & i\,P_0(\vec{\boldsymbol{\xi}})\,\phi_0(\vec{\boldsymbol{\xi}})\\
    \mathrm{Var}\left[<\Psi_S(\vec{\boldsymbol{\xi}},\,t)>_{\tilde{N}}\right] &\displaystyle  P_0(\vec{\boldsymbol{\xi}})\,\frac{\sigma_1^2}{\tilde{N}}.
\end{array}
\right.    \label{eq:psi_s_fin_var_small}
\end{equation*}
In other words, the averaged electric field at the Lyot plane is
\begin{equation}
    <\Psi_S(\vec{\boldsymbol{\xi}},\,t)>_{\tilde{N}} = i\,P_0(\vec{\boldsymbol{\xi}})\,\left[\phi_0(\vec{\boldsymbol{\xi}})\pm \epsilon_{\tilde{N}}~(\vec{\boldsymbol{\xi}})\right],
    \label{eq:estimate_phi}
\end{equation}
where~$\epsilon_{\tilde{N}}(\vec{\boldsymbol{\xi}})$ is a complex random function that follows a Gaussian distribution of variance~$\sigma_1^2/\tilde{N}$. If a FPWFS can measure $<\Psi_S(\vec{\boldsymbol{\xi}},\,t)>_{\tilde{N}}$, the averaged phase~$\phi_0+<\phi_1>_{\tilde{N}}$ can be estimated directly from Eq.~\ref{eq:estimate_phi}. The longer the exposure, the more accurate the estimation of $\phi_0$.

\section{Numerical simulations}
\label{s:resultNum}
\begin{figure}
     \includegraphics[width=7cm]{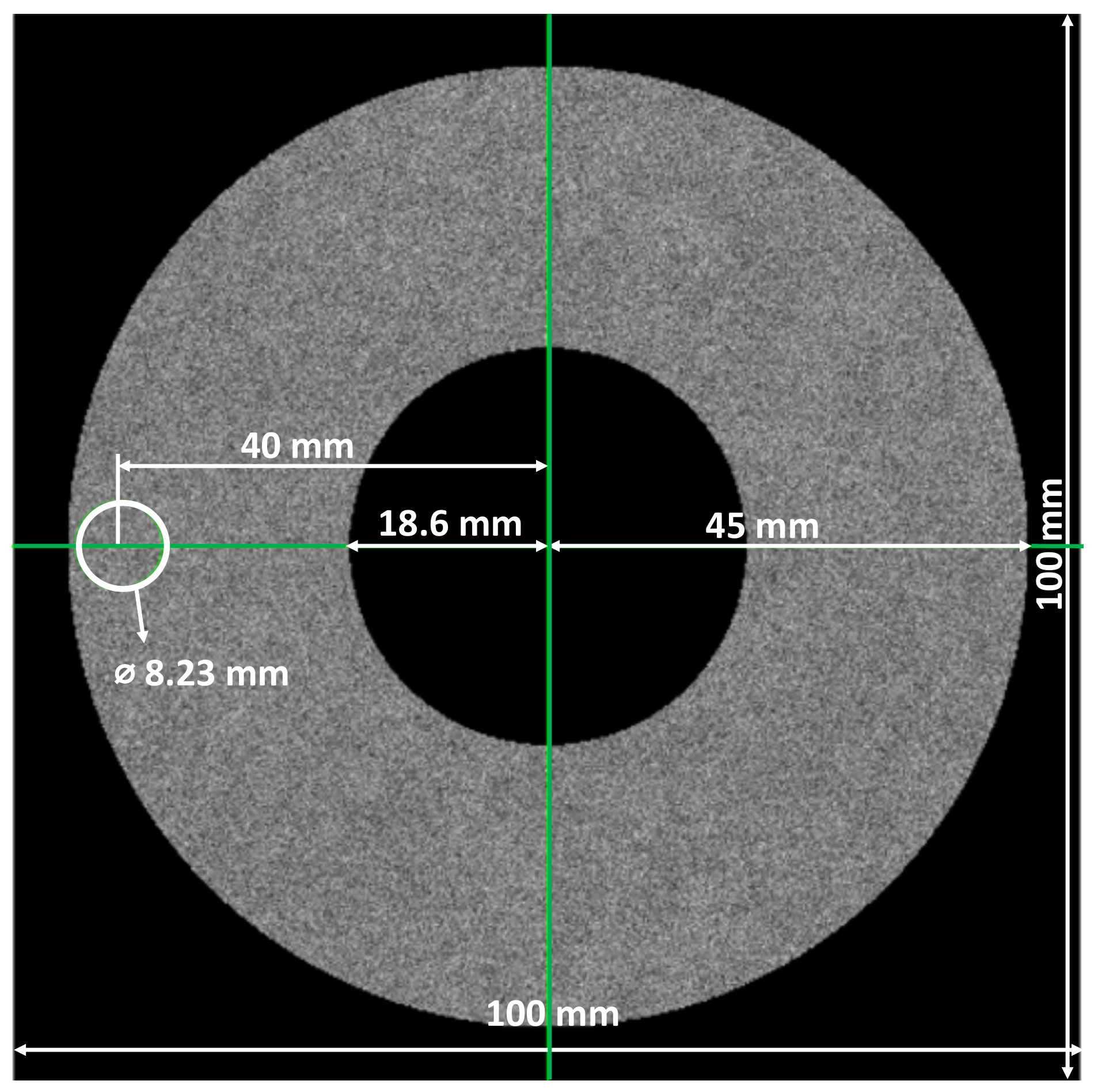}
      \caption{Simulated map of the phase function that is induced by the phase plate. The entrance pupil of the THD2 bench is represented by a white circle. The phase plate is rotated around its center to mimic post-AO residual turbulence.}
         \label{fig:turb_map}
   \end{figure}
As stated above, the goal is to estimate the static aberrations~$\phi_0$ in long exposure in the presence of evolving aberrations~$\phi_1$. We assume here that a system is able to estimate~$\psi_S$ that averages as described in section~\ref{s:wave}. In the following sections, we first simulate the post-AO residuals using the assumptions that reproduce the laboratory conditions presented in  section~\ref{s:thd2}. We then determine what level of static aberrations can be measured for a given exposure time and what level of contrast is achievable in the corresponding coronagraphic images. 

\subsection{Assumptions}
\label{s:resultNum1}

For the experimental demonstration, we optically simulate the post-AO residuals using a rotating phase plate on the THD2 bench (more details about the practical implementation are given in section~\ref{s:thd2}). The simulated version of the phase plate is shown in Fig.~\ref{fig:turb_map}. In the coronagraphic system as shown in Fig~\ref{fig:coro_layout_1}, we consider the entrance pupil to be $8.23$\,mm in diameter and to be located at $40$\,mm from the center of the rotating phase plate. The plate can rotate around its center with $12,000$ steps in total at a speed set to $300$ steps per second. This corresponds to a linear velocity of $6.28$\,mm/s, or $6.10$\,m/s when scaled to an 8m telescope. The power spectral density (PSD) of the simulated aberrations in Fig.~\ref{fig:turb_map} is similar to what is measured behind the AO system of  SPHERE~\citep{fusco06a} in the infrared. The PSD follows a power law as $f^{-4/3}$ for spatial frequencies lower than 20 cycles per pupil and as $f^{-11/3}$ for higher spatial frequencies. The standard deviation of the aberration is assumed to be~$40$\,nm (in visible) inside the pupil mainly distributed in spatial frequencies below 2\,cycles per pupil diameter and around the cut-off of~20\,cycles per pupil diameter. Though such a level of aberration is optimistic for the current version of SPHERE (Beuzit et al. 2019, submitted), it represents a good estimation of the performance of an upgraded SPHERE.

\begin{figure}
     \includegraphics[width=9cm]{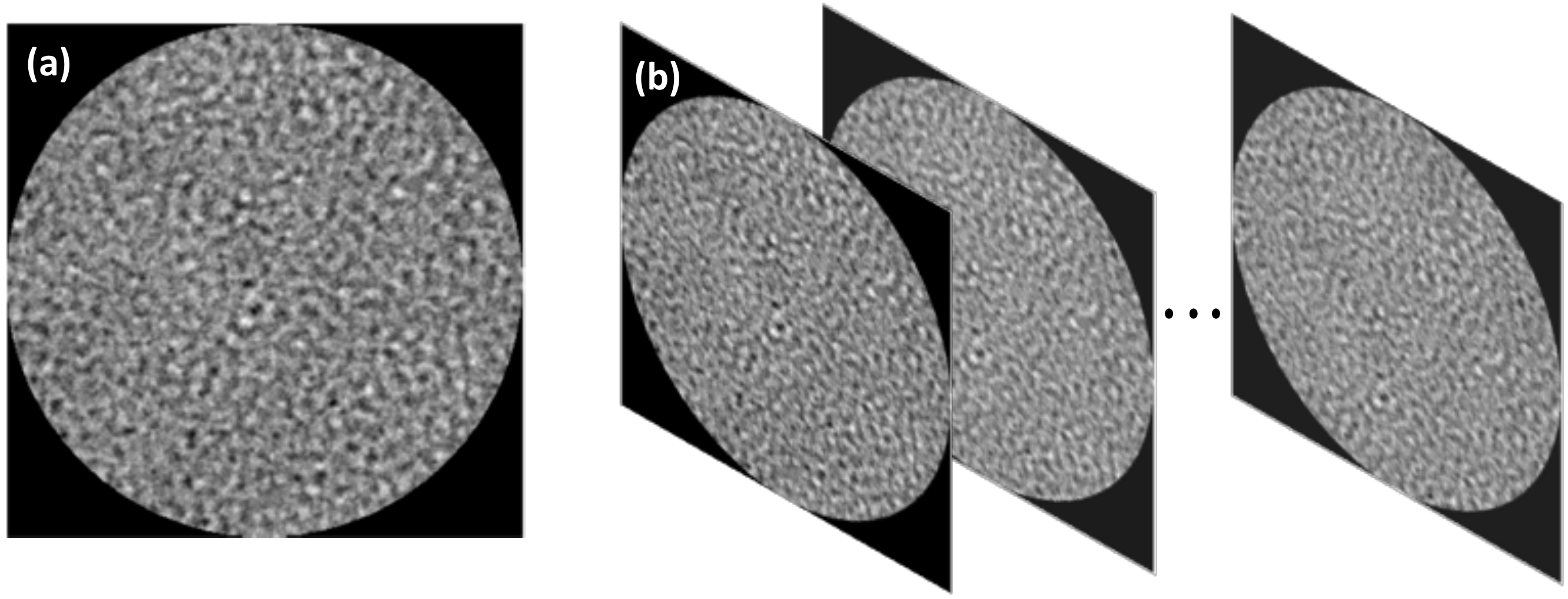}
      \caption{(a) Snapshot of one individual phase screen seen through the THD2 pupil. (b) A cube of 12,000 phase screens used in the simulation.}
         \label{fig:phases}
   \end{figure}

We simulate the rotation of the post-AO phase plate around its center and extract $12,000$ individual phase screens~$\phi_1(t)$ generated by the plate in the entrance pupil of $8.23$\,mm as shown in Fig.~\ref{fig:turb_map}. These extracted phase screens are shown in Fig.~\ref{fig:phases}.
   
\subsection{Accuracy versus exposure time}
\label{s:resultNum2}
\begin{figure}
     \includegraphics[width=9cm]{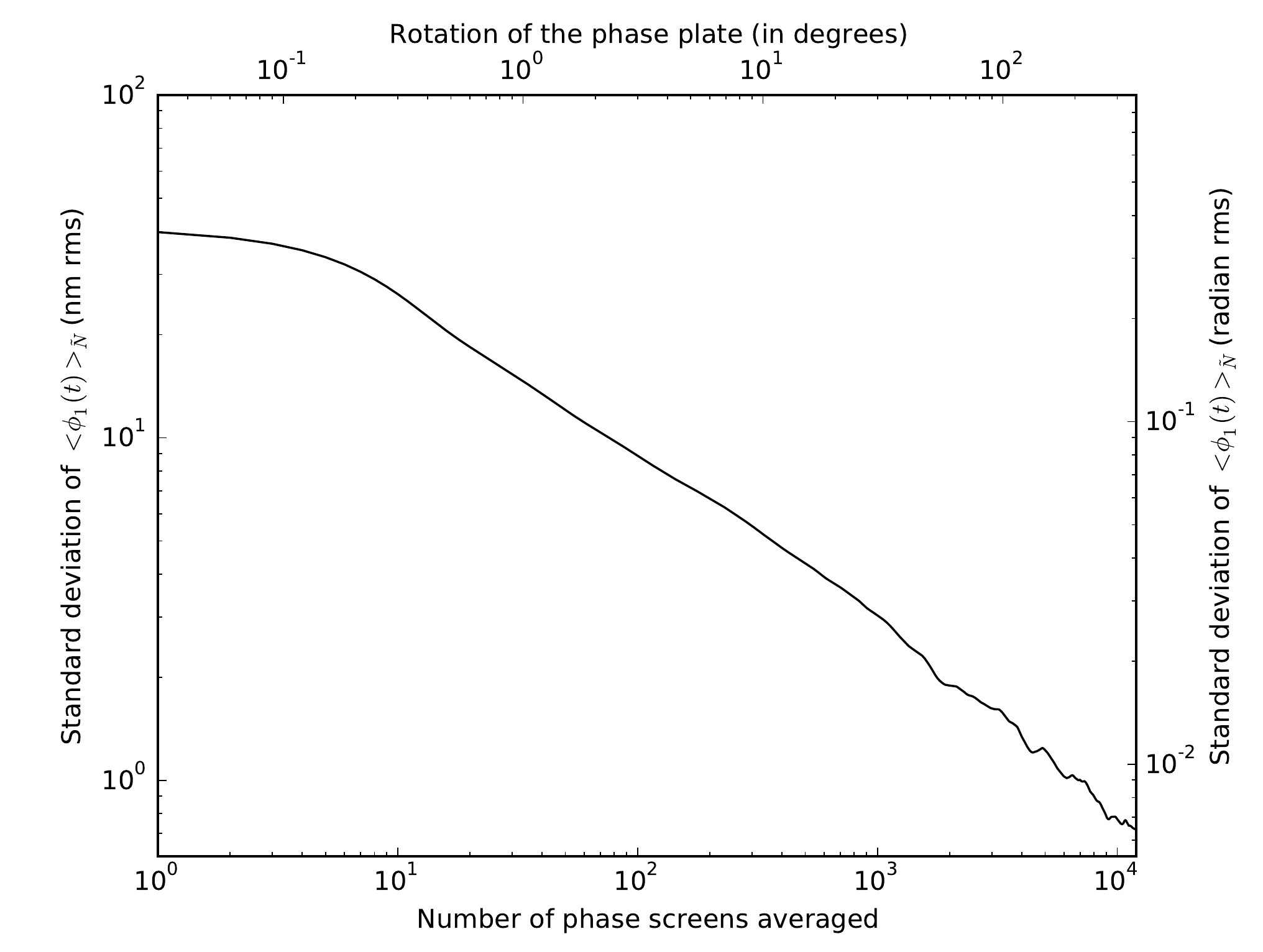}
      \caption{Simulated standard deviation within the pupil of the phase aberration averaged over~$\tilde{N}$ individual phase screens. The upper x-axis reads the angle of rotation of the phase plate during the exposure. The lower x-axis presents the average number of phase screens~$(\tilde{N})$ seen by the pupil. The y-axes shows the standard deviation of the averaged phase inside the pupil in nm rms (at~$700$\,nm) on the left and in radian rms on the right.}
         \label{fig:rms}
   \end{figure}

In the next step, we calculate the average of the phase screens as ~$<\phi_1(t)>_{\tilde{N}}$ with~$\tilde{N}$ increasing from $1$ to~$12,000$. Figure~\ref{fig:rms} presents the standard deviation of the averaged phases~$<\phi_1(t)>_{\tilde{N}}$ as a function of~$\tilde{N}$.
The plotted data give the level of aberrations for a finite long exposure in the absence of static aberrations. In other words, whatever the FPWFS, the best accuracy on the measured static aberrations from a finite long exposure is given by~Fig.~\ref{fig:rms} for the simulated AO system. For example, for a plate rotation of $162^\circ$ (see section~\ref{s:resultExp}), the entrance pupil sees $5,400$ simulated phase screens ($\tilde{N}=\frac{12,000\times 162}{360}$) pass by, and the standard deviation of the post-AO residuals is~$\sim1\,$nm. It is therefore expected that a FPWFS on the assumed experimental bench (section~\ref{s:thd2}) should be able to measure aberrations with an accuracy of~$\sim1\,$nm for an exposure time corresponding to a $162^\circ$ rotation of the phase plate.

More generally, we find that the standard deviation of the averaged phases $<\phi_1(t)>_{\tilde{N}}$ decreases as the square root of the number of individual screens considered in the range~$150\lesssim \tilde{N}\lesssim1,000$. For $\tilde{N}\lesssim150$, the standard deviation decreases at a rate slower than~$\sqrt{\tilde{N}}$ because the phase screens are not completely uncorrelated. There are mainly two regimes for $\tilde{N}\lesssim150$: residuals decreasing very slowly for~$\tilde{N}\lesssim10$ and a bit faster for~$\tilde{N}>10$. The pupil beam of diameter $8.23$\,mm hits the simulated post-AO phase plate at~$40$\,mm from its center as shown in Fig.~\ref{fig:turb_map}. For $\tilde{N}\lesssim10$, the rotation of the post-AO phase plate is less than~$\frac{2\,\pi\,40\times10}{12,000}\sim0.21\,$mm, which is equivalent to a movement of the plate by less than~$\sim1/40$ of the pupil diameter. From the design of the phase plate (section~\ref{s:thd2}), most of the energy in the PSD of~$\phi_1$ is close to the cut-off at $20$\,cycles per pupil. It is therefore expected that the standard deviation of~$<\phi_1(t)>_{\tilde{N}}$ decreases at a rate slower for~$\tilde{N}\lesssim10$ than for~$\tilde{N}>10$. Furthermore,~$\tilde{N}=150$ corresponds to a spatial frequency of two cycles per pupil, which is where the rest of the energy lies in the PSD of~$\phi_1$. For~$\tilde{N}>1,000$, we observe variations because the variance of the two independent phase screens is not exactly the same.

\subsection{Coronagraphic images}
\label{s:resultNum3}

\begin{figure*}
              \centering  
     \includegraphics[width=15cm]{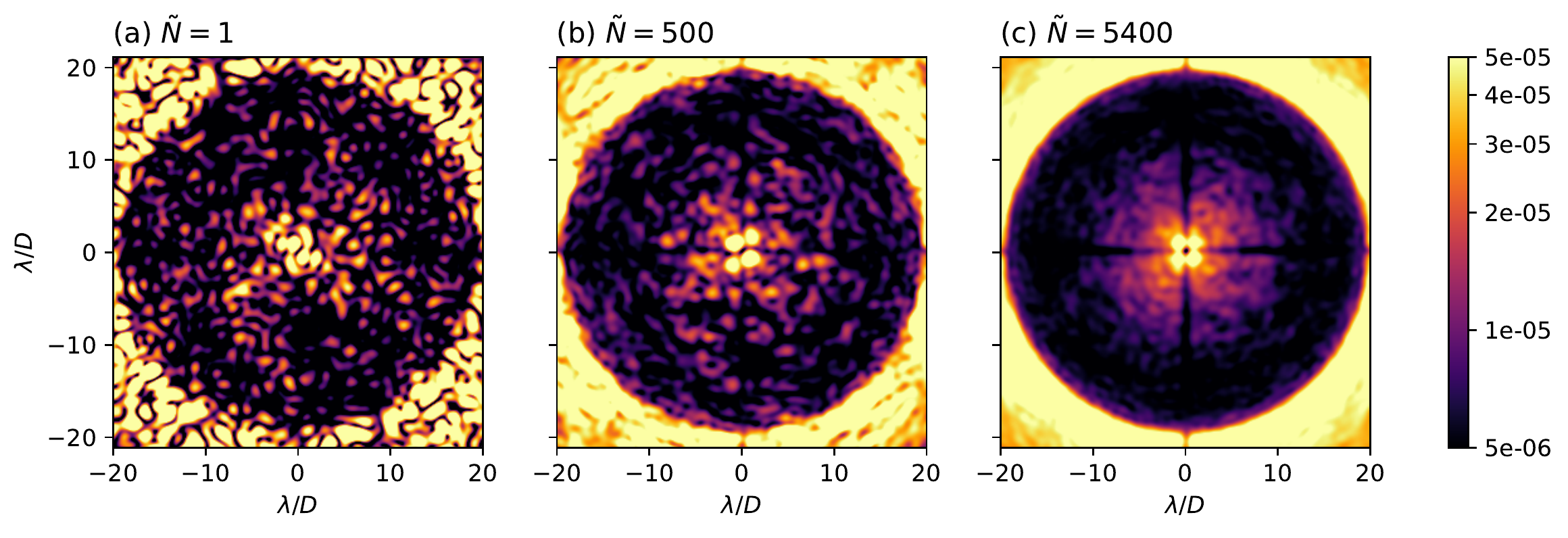}
      \caption{Focal plane coronagraphic images simulated with the FQPM coronagraph under the post-AO wavefront errors representing the residual level seen by SPHERE when atmospheric  conditions are favorable. No additional static aberrations are added. (a)~Short exposure freezing the AO wavefront ($\tilde{N}$ = 1) at $ 783.25$\,nm. The radius of the AO-correction cut-off is 20~$\lambda/D$. The speckles smoothed out when (b)~$\tilde{N}$ = 500 and (c)~$\tilde{N}$ = 5400 short-exposure images are averaged. With increasing $\tilde{N}$, a gradual decrease of speckles can be observed in the images. The dark cross in all the images is due to the FQPM transitions. All images are at the same brightness scale.}
         \label{fig:sim_images}
   \end{figure*}

To simulate long-exposure images under the post-AO residuals, we compute one coronagraphic image per individual phase screen (frozen aberrations) following the optical layout of Fig.~\ref{fig:coro_layout_1}. We assume a monochromatic light at~$783.25$\,nm at the entrance pupil and a four-quadrant phase mask \citep[FQPM,][]{rouan} as a focal plane mask (FPM). Only the post-AO aberrations without any static errors are considered ($\phi_0=0$). The intensities of the~$\tilde{N}$ individual images are then averaged to obtain a long-exposure coronagraphic image. Figure~\ref{fig:sim_images}~(a) represents a science image obtained for a short exposure ($\tilde{N}=1$) where AO residuals are frozen.
 Figures~\ref{fig:sim_images}~(b) and~(c) show how the AO halo averages when the exposure time ($\tilde{N}=500$ and $5,400$) is increased. The darker area in all the images represents a region controlled by the simulated AO system. The cut-off is at~$20\,\lambda/D$ (40 actuators across the pupil in the SPHERE instrument). We notice that even for $\tilde{N}=5,400$ averaged images, speckles are still visible in the controlled area. They are induced by the~$\sim1\,$nm rms residual aberrations found in~Fig.~\ref{fig:rms} and they vary from one long exposure to another. The dark cross that is visible in the images is due to the transitions of the FQPM. All the images are normalized by the maximum of the non-coronagraphic image.     

\begin{figure}
     \includegraphics[width=9cm]{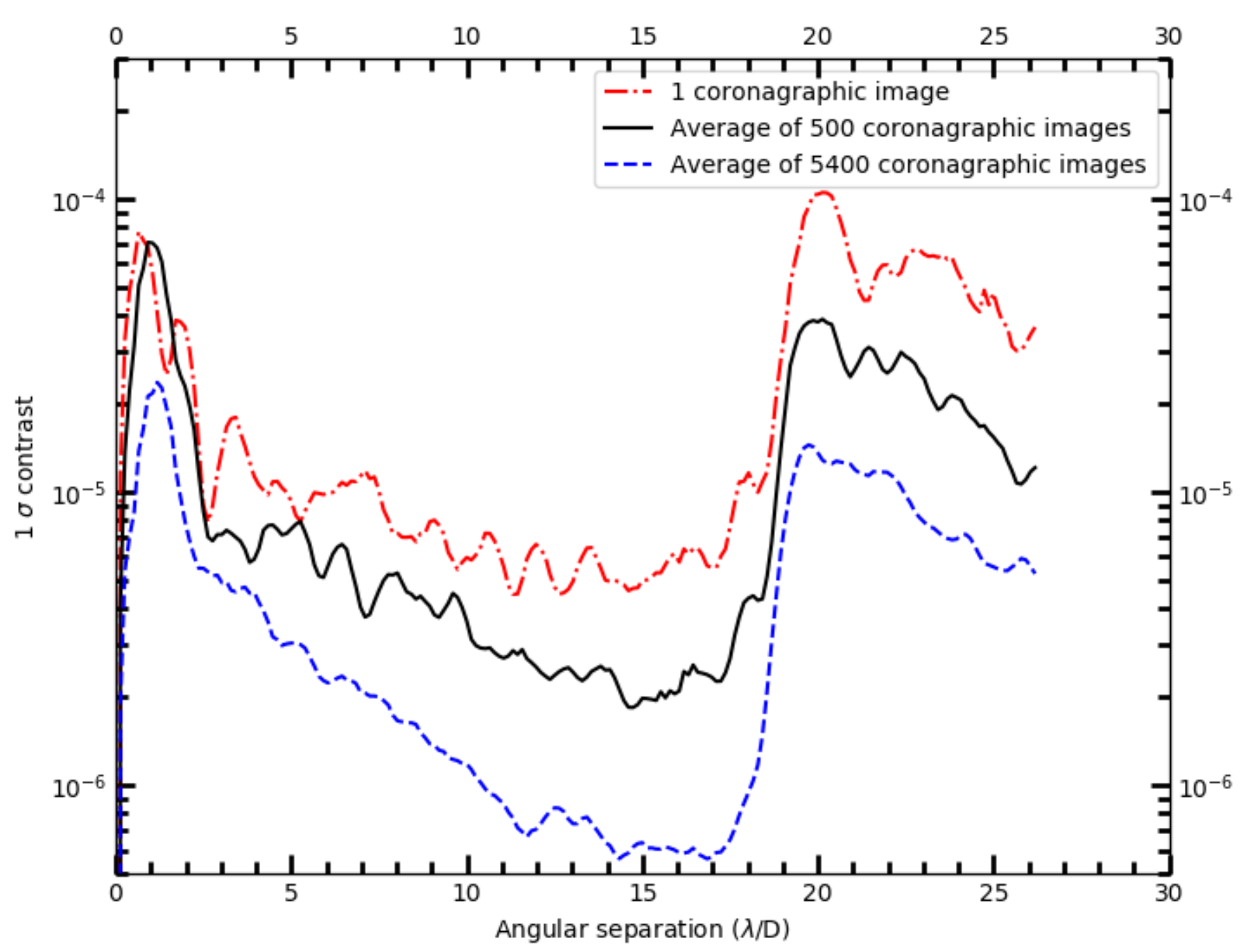}
      \caption{Simulated profiles showing azimuthal standard deviation as a function of angular separation for focal plane images presented in Fig.~\ref{fig:sim_images}. The red profile shows the contrast curve of a short-exposure image depicting what is obtained by SPHERE under good observing conditions. The contrast in the range $5-18\,\lambda/D$ mostly stays around 10$^{-5}$. For a realistic long exposure, the blue profile is mostly below 3$\times$10$^{-6}$ in the range $5-18\,\lambda/D$. }
         \label{fig:sim_contrast}
   \end{figure}

Figure~\ref{fig:sim_contrast} represents the azimuthal standard deviation profiles ($1\,\sigma$ contrast) associated with the images of Fig.~\ref{fig:sim_images}.
At small angular separations ($<\,3\,\lambda/D$), starlight leaks due to low spatial frequency aberrations are visible. When speckles are frozen (red curve), the raw contrast reaches the level of $\sim10^{-5}$ between~$5$ and $18\,\lambda/D$. This contrast roughly corresponds to what is measured behind the SPHERE AO system at the VLT \citep{vigan2019}. For longer exposure, the quickly varying speckles average out and the contrast deepens. By choosing a realistic exposure time ($N=5,400$, which also corresponds to $18\,$s exposure at the VLT), the contrast level reaches below~$3\times10^{-6}$ in the range $5-18\,\lambda/D$ (blue curve). Therefore, for an exposure of $18\,$s under the simulated AO residuals, the contrast of $\sim\,6\times10^{-7}$ is the best level that can be reached after the expected correction of quasi-static aberrations by any FPWFS. 
   
\section{THD2 bench in a glance}
\label{s:thd2}
For our practical experiments, we used the THD2 bench. The detailed description of THD2 optical components can be found in \citet{baudoz2018}. In this section, we briefly recall the main components of THD2 and a sub-system that optically simulates the post-AO residuals. With a growing collaboration all over the world, the THD2 bench became a unique research and development platform for HCI in Europe. The main objective is to optimize the design of future exoplanet imaging instruments by comparing several HCI techniques under the same environmental conditions. Though the THD2 bench is not in a vacuum tank, the stabilization of temperature and humidity already enabled the testing of HCI techniques under steady conditions similar to those in space. 

\begin{figure*}
          \centering  
   \includegraphics[width=14cm]{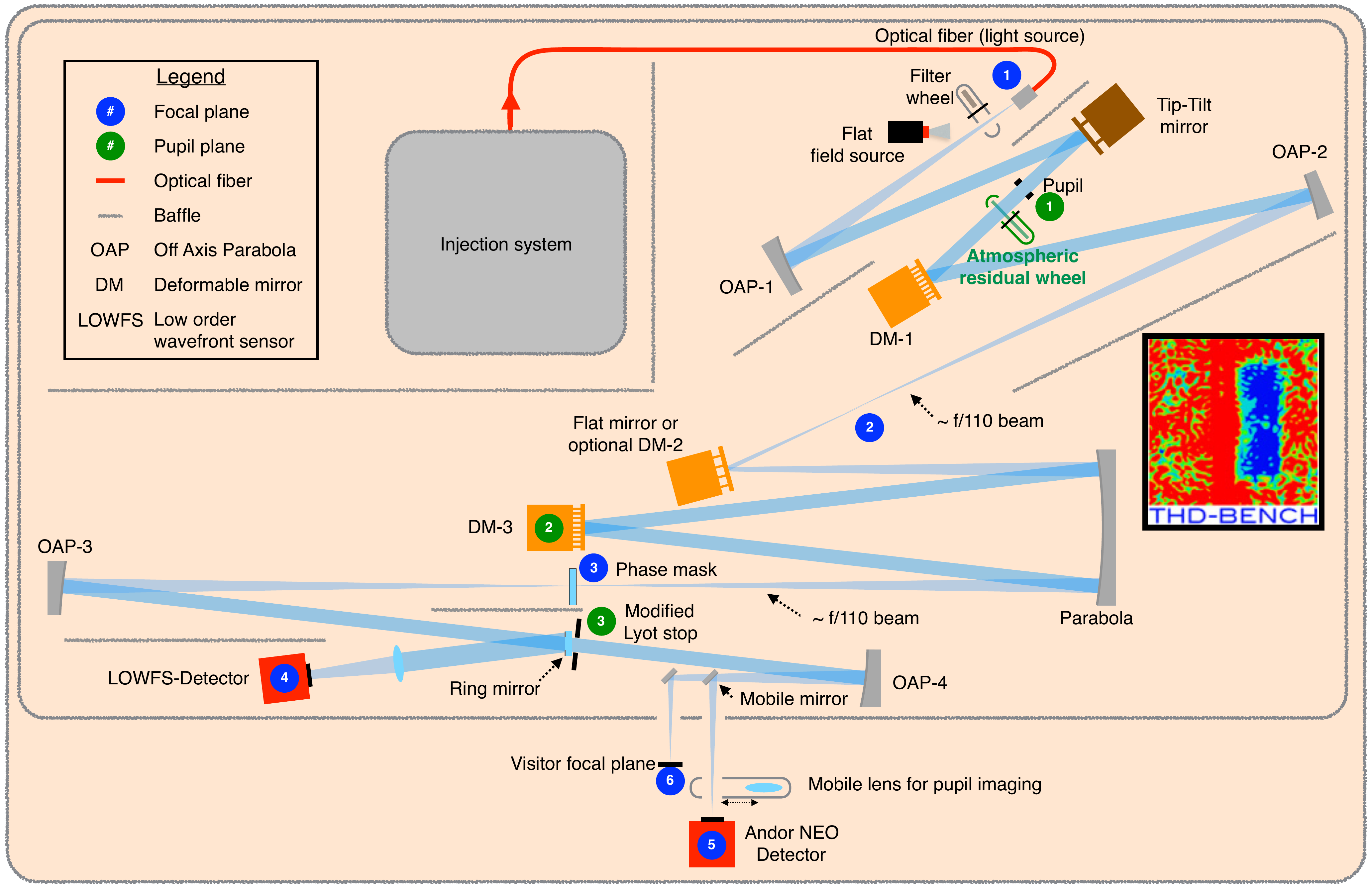}
      \caption{Optical diagram of the THD2 bench. The post-AO residual subsystem containing a rotating phase plate is situated immediately after the entrance pupil.}
         \label{fig:thd}
   \end{figure*}

\begin{figure}
   \includegraphics[width=9cm]{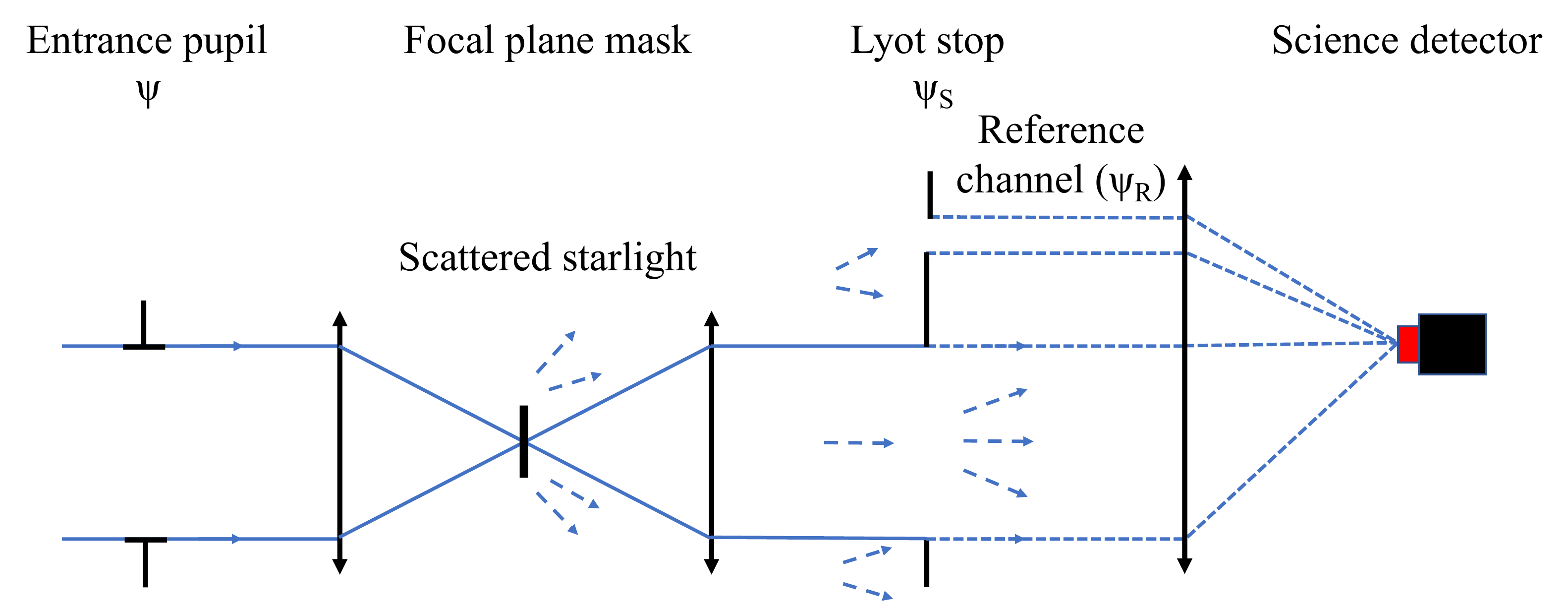}
      \caption{Optical layout of the SCC as used on the THD2 bench. The Lyot stop has a classical on-axis diaphragm and a reference pupil for the SCC. The aberrated starlight that goes through the on-axis diaphragm combines with the reference starlight to spatially modulate the speckle intensity in the science image.}
         \label{fig:coro_layout}
   \end{figure}

The optical representation of the THD2 bench is shown in Fig.~\ref{fig:thd}, which follows the coronagraphic layout of Fig.~\ref{fig:coro_layout}.  
   \begin{figure}
   \includegraphics[width=9cm]{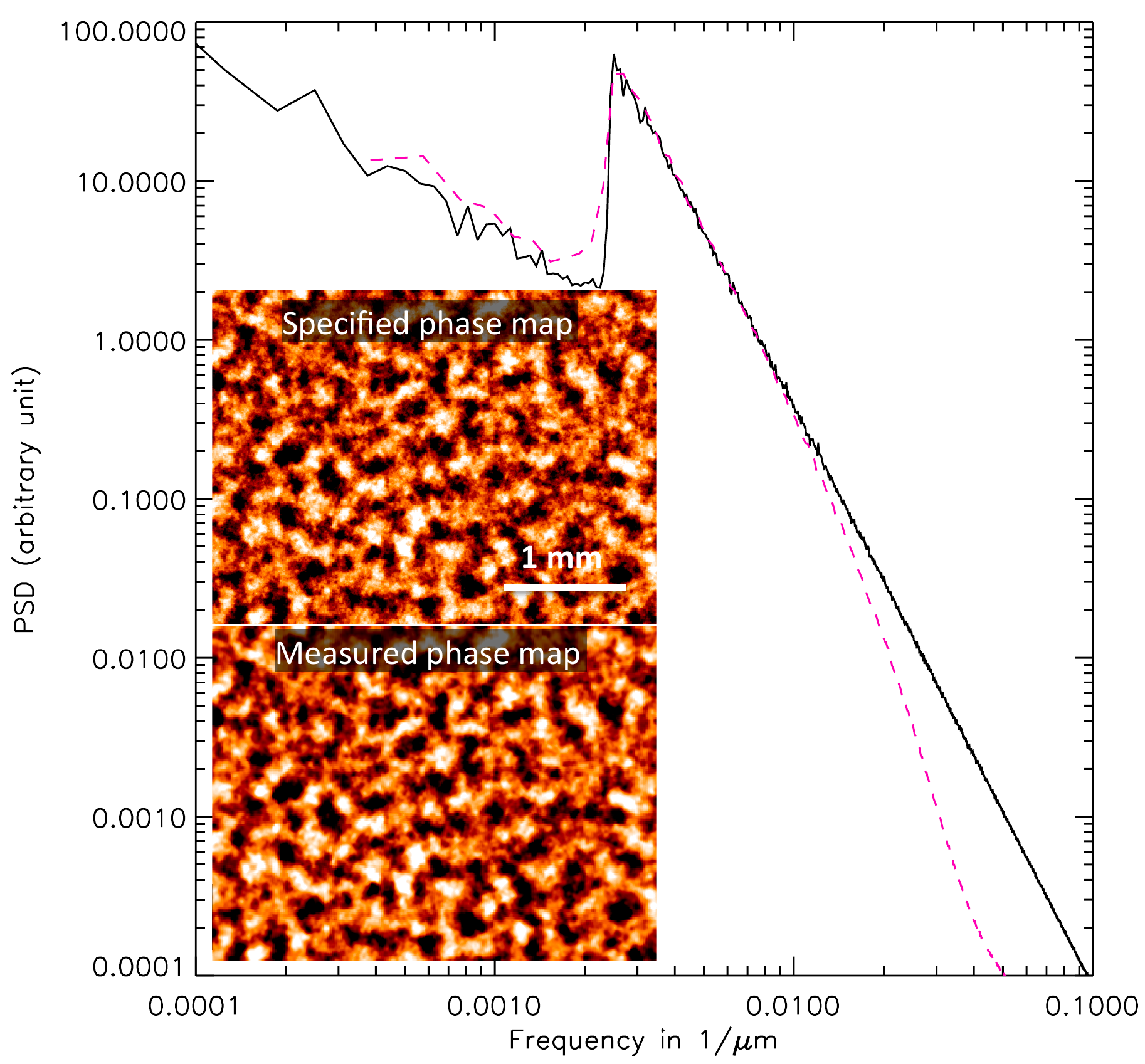}
      \caption{Theoritical (black profile) and measured (red dashed line) PSD of the phase plate shown in Fig.~\ref{fig:phases}. The images of theoretical and measured phase maps are extracted  within the area of $3.49\,mm\times2.62\,mm$ on the phase plate.}
         \label{fig:psd1}
   \end{figure}
   To simulate starlight, a  single-mode optical fiber injects light at the entrance of the bench. In this paper, the source is a laser diode emitting at $\lambda = 783.25$\,nm. The starlight is collimated towards a tip-tilt mirror (TTM), which is used to control the pointing. The beam after the correction of tip-tilt errors meets a circular unobscured entrance pupil, the diameter of which is set to $8.23$\,mm in this paper. The beam is then reflected by two DMs from Boston Micromachine Corporation (BMC). The first, DM1, has 952 actuators and is set in a collimated beam at~$26.9\,cm$ from the pupil plane. The second, DM3, has 1024 actuators and is conjugated to the pupil plane. There are~$27$ actuators of each DM across the pupil. After~DM3, the beam is focused onto a transmissive FPM with a f-number of~$110$. In this paper, the FPM is a FQPM coronagraph. In the following pupil plane, a Lyot stop filters the starlight that is scattered by the FPM outside the geometrical pupil. The diameter of the Lyot diaphragm $D_L$ is $8.00\,$mm for our experiments. An off-axis hole of~$0.5$\,mm in diameter is also used to create the SCC reference channel (section~\ref{s:scc}). Finally, the coronagraphic image is recorded by an Andor camera with a readout noise of 3.2~$e^{-}$~rms per pixel. The science images are 400$\times$400 pixels with the resolution element sampled by 7.6 pixels ($\lambda/D_{L}$). 

To prepare new ground-based imaging instruments or propose upgrades to the existing ones, an optical subsystem has been installed on THD2 to mimic post-AO residual aberrations in visible. The simulated post-AO phase plate is shown in Fig.~\ref{fig:turb_map} and its specifications are explained briefly in Sect.~\ref{s:resultNum}. It is composed of a transmissive ($>99\%$ between 600 and 800~\,nm) rotating phase plate with a diameter  of 100\,mm and a thickness of~$1.5$\,mm. The aberrations introduced on THD2 have a PSD following a power law as $f^{-4/3}$ and $f^{-11/3}$ for lower ($<$20 cycles per pupil) and higher spatial frequencies, respectively. The specified standard deviation of the aberration is~$40$\,nm inside the pupil. The plate was fabricated by the Zeiss company using ion-etching. The etched part of the plate starts at 18.6mm from its center and continues until 45mm. The lateral resolution of the etched phase function is~$10\mu$m. Figure~\ref{fig:psd1} presents the curves of a theoretical and measured PSD of the phase plate. The measured PSD is calculated using the images provided by an interferometric microscope with a field of view of $ 6.98\,mm\times5.24\,mm$ and a pixel size of $0.64\mu$m. These images are recorded on eight areas evenly distributed on the edge of the phase plate. Both a theoretical and measured phase map for a small area of $3.49\,mm\times2.62\,mm$ are also shown in Fig.~\ref{fig:psd1}. The plate sits immediately after the entrance pupil (green component in~Fig.~\ref{fig:thd}). It can rotate around its center, thus continuously changing the phase errors in the 8.23mm entrance pupil (white circle in~Fig.~\ref{fig:turb_map}). The encoder that controls the position of the plate uses 12,000 steps for a full round. In the experiments presented here, the plate rotates by 300 steps per second and the beam that goes through the pupil hits the phase plate at~40mm from its center.

\section{Focal-plane wavefront estimation in long exposure}
\label{s:scc} 
 The two main assets of THD2 bench for wavefront measurement are the Lyot-stop low-order wavefront sensor \citep[LLOWFS,][]{singh3} and a FPWFS. The former is used to stabilize the tip-tilt at~$<100$~Hz and the latter stabilizes all the other spatial frequencies up to the DM cut-off ($13.5\,\lambda/D$). As for the focal plane wavefront sensing and correction, we can either use pair-wise plus electric field conjugation (Potier et al. in Prep) or the SCC. In the experiments presented in section~\ref{s:resultExp}, we used both the LLOWFS and the SCC to control the wavefront errors.
 
The SCC principle has already been described in several papers~\citep{scc2,scc5}. The SCC uses a modified Lyot stop as shown in~Fig.~\ref{fig:coro_layout}. The on-axis diaphragm is the classical Lyot stop that filters out the stellar light rejected by the coronagraphic FPM. Part of the starlight goes through the diaphragm because of aberrations, which induce speckles on the science detector. We use~$A_S$ to refer to the electric field of these speckles. The off-axis reference hole selects part of the rejected starlight at a distance~$\vec{\boldsymbol{\xi_0}}$ from the center of the on-axis diaphragm. Doing so, the stellar speckles are spatially modulated by Fizeau fringes on the detector. If we consider that the aberrations in the system are static during the exposure time, the intensity $I(\vec{x})$ at a position~$\vec{x}$ in the science image can be written as
\begin{equation}
    I(\vec{x}) = \left|A_S(\vec{x})\right|^2 + \left|A_R(\vec{x})\right|^2 + 2\mathcal{R}\left(A_R^*(\vec{x})\,A_S(\vec{x})\,e^{\displaystyle\frac{2i\pi\vec{x}.\vec{\boldsymbol{\xi_0}}}{\lambda}}\right),
\label{eq:iscc}
\end{equation}
where~$A_R^*$ is the conjugate of the electric field in the science image associated with the off-axis hole of the Lyot stop and, $\mathcal{R}$ is the real part of a complex number. It has been demonstrated that~$I_-=A_R^*\,A_S$ can be extracted from~$I$ and that minimizing~$I_-$ also minimizes the speckle electric field~$A_S$ in the controlled area. To estimate~$I_-$, one selects the lateral peak in the Fourier transform of~Eq.\,\ref{eq:iscc}. The expression of this peak is
\begin{equation}
\hat{I}_- = \Psi_S*\Psi_R^*,
\label{eq:iscc_im}
\end{equation}
where~$\Psi_R^*$ is the conjugate of the electric field in the off-axis hole of the Lyot stop, ~$\Psi_S$ is the field inside the on-axis diaphragm, and $*$ denotes the convolution product. If the wavefront aberrations are not too large or the diameter of the off-axis hole is small enough, it can be assumed that $\Psi_R$ does not depend on the wavefront aberrations \citep{scc2,sc}. On the contrary, $\Psi_S$ strongly depends on the aberrations upstream of the coronagraph. As shown in section~\ref{s:wave}, the SCC can retrieve the static aberrations~$\phi_0$ from the estimation of $<\Psi_S(\vec{\boldsymbol{\xi}},\,t)>_{\tilde{N}}$ measured from a long exposure coronagraphic image. 

\section{Laboratory performance}
\label{s:resultExp}
\begin{figure*}
          \centering  
   \includegraphics[width=13cm]{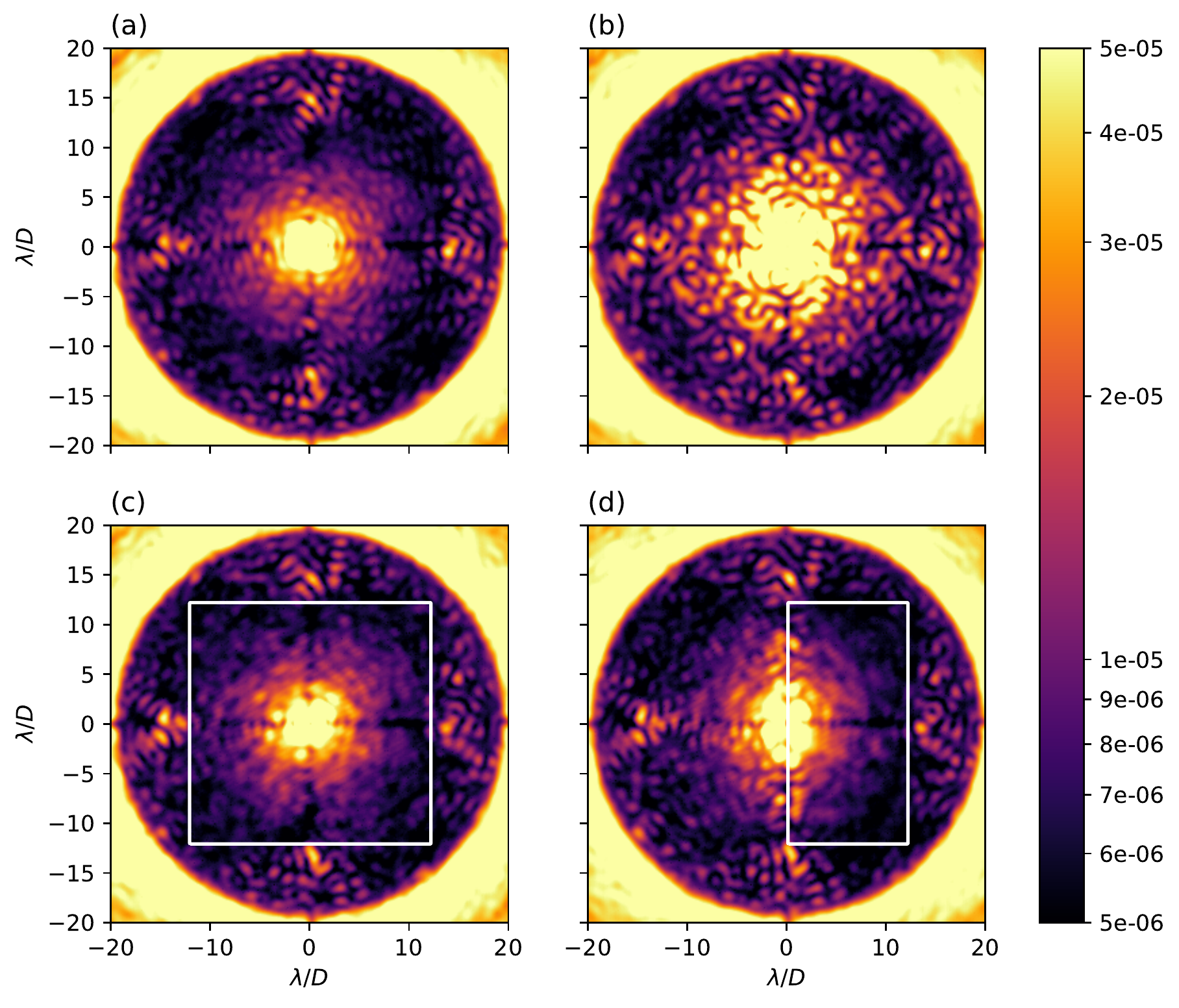}
      \caption{Laboratory SCC science images obtained at an exposure of 18\,s on the THD2 bench: (a)~under the effect of post-AO aberrations only and (b, c, d) under the effect of both post-AO aberrations and static aberrations. (b)~~$0^{th}$ iteration of the~SCC correction loop (starting point). (c) and (d)~are obtained after five iterations of the~SCC correction loop when controlling speckles inside a full dark hole within a~$25\,\lambda/D\times25\,\lambda/D$ region and a half dark hole going from~$-12.5\,\lambda/D$ to~$12.5\,\lambda/D$ in one direction and from~$2\,\lambda/D$ to~$12.5\,\lambda/D$ in the other direction.}
         \label{fig:lab_images}
   \end{figure*}

\begin{figure*}
              \centering       \includegraphics[width=16cm]{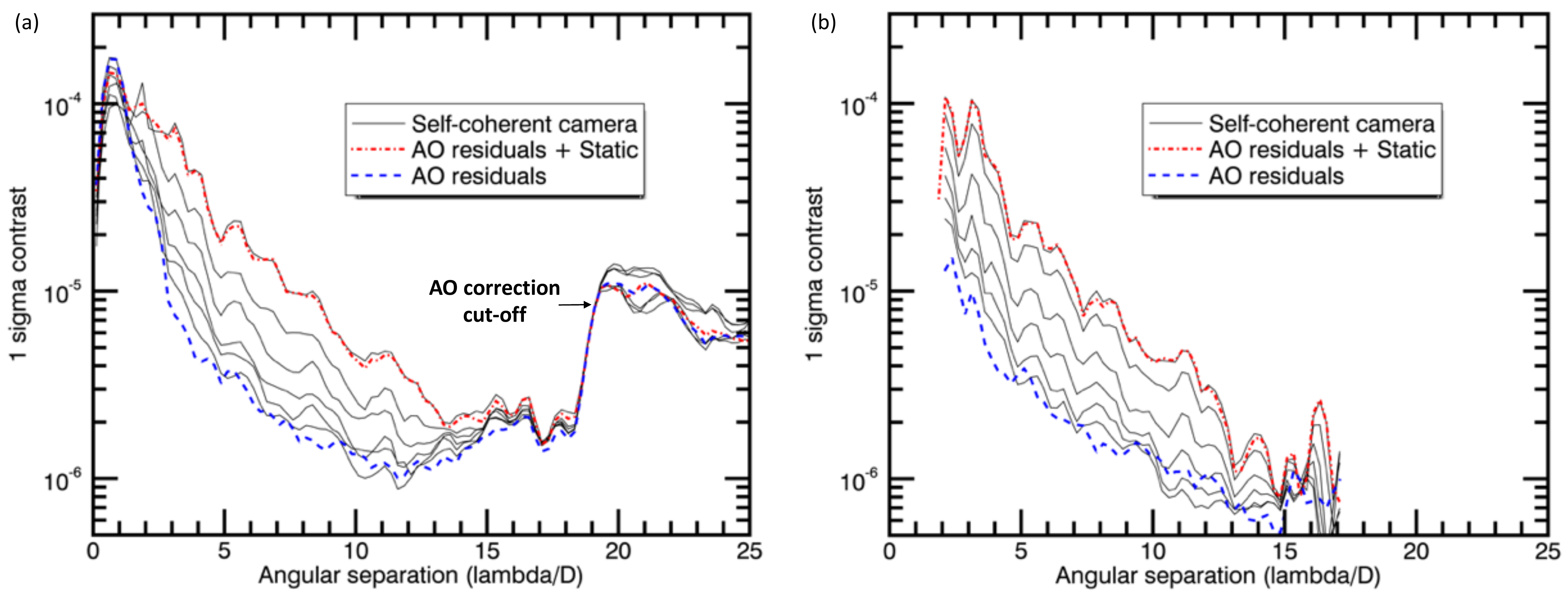}
      \caption{Azimuthal standard deviations (i.e., $1\,\sigma$ contrast detection) of the~18\,s exposure images of~Fig.~\ref{fig:lab_images}. The blue dashed curves represent the post-AO aberrations if no static aberrations are added. The red dot-dash ones show the post-AO aberrations after adding static aberrations. The black profiles from top to bottom show the first six iterations of correction in the case of (a)~the full dark hole and (b)~the half dark hole. The $0^{th}$ iteration in a black curve is similar to the post-AO residuals with static aberrations.}
         \label{fig:lab_contrast}
   \end{figure*}

In this section, we demonstrate that in the laboratory the SCC can be used to compensate static errors down to the fundamental level set by the averaged AO residuals described in section~\ref{s:resultNum}. Before beginning our experiments, we first minimized the speckle level at $<10^{-8}$ contrast using the SCC with no AO phase plate in the beam. We then recorded a series of long-exposure SCC images by choosing a realistic exposure of~$18$\,s while the phase plate that optically simulates the optimistic level of SPHERE post-AO residuals rotated at~$300$\,steps per second. One such image is shown in Fig.~\ref{fig:lab_images}~(a). It sets the fundamental level of AO residuals for~$18$\,s exposure for the simulated AO system. The corresponding~$1\,\sigma$ contrast curve is plotted in blue (dashed line) in~Fig.~\ref{fig:lab_contrast}.
  Figure~\ref{fig:lab_images}~(a) is similar to the numerically predicted coronagraphic image shown in Fig.~\ref{fig:sim_images}~(c). The simulated AO cut-off is at~$20\,\lambda/D$ and the distribution of energy (blue dashed curves) as a function of spatial frequencies in~Fig.~\ref{fig:lab_contrast} is also similar to the blue curve in~Fig.~\ref{fig:sim_contrast}.
There are two small differences however: First, the laboratory image is brighter close to the star center. Even though the tip-tilt loop was closed, additional low-order aberrations were introduced by the plate itself that added up with the etched errors. Second, there are a few speckles outside the $13\,\lambda/D\times13\,\lambda/D$ area, which are brighter in the laboratory image. They are uncontrolled static aberrations on the THD2 bench and cannot be compensated as the cut-off of THD2 DMs is~$13\,\lambda/D$.

In the next step, both phase and amplitude static aberrations were added to the system using the~DM1 (inducing~$1$\,nm~rms of phase and~$0.4\,\%$ of amplitude) and~DM3 (inducing~$5$\,nm~rms of phase only). The spectral density function of these aberrations varies as the inverse of the spatial frequency ($f^{-1}$). This level of static aberrations is similar to what SPHERE can achieve after compensating the NCPA on the internal source. After applying these static aberrations in our system, we let the post-AO phase plate rotate. In the mean time, the science camera continuously recorded images, each acquired with an exposure of~18\,s. One such aberrated image of~$18\,$s exposure is presented in~Fig.~\ref{fig:lab_images}~(b). The associated contrast curve is plotted in red dot-dash in~Fig.~\ref{fig:lab_contrast}. As expected, static speckles limit the contrast performance. At each exposure, the~SCC was then used to estimate and control the aberrations by using DM3 which is conjugated to the pupil plane. We note that the interaction matrix of the SCC was recorded under no post-AO residuals, which is similar to what would be done at the telescope using the internal source of the instrument. 

In the first test, we used the control algorithm to minimize the speckle intensity in a full dark hole of~$25\,\lambda/D\times25\,\lambda/D$ centered on the star. Figure~\ref{fig:lab_images}~(c) shows the long-exposure image recorded by the camera after five iterations of correction (i.e., six images). The contrast curves for the first five iterations of the~SCC correction are presented in~Fig~\ref{fig:lab_contrast}~(a) in black. The static speckles were corrected and the science image is at the level of the post-AO residuals (Fig.~\ref{fig:lab_images}~(a)) that no FPWFS can overcome for a given exposure time~(see section~\ref{s:resultNum}). The only difference between Fig.~\ref{fig:lab_images}~(a) and~\ref{fig:lab_images}~(c) is that they were acquired at different rotation angles of the phase plate.

In the second test, we started again from the image with static aberrations shown in~Fig.~\ref{fig:lab_images}~(b). We then commanded the algorithm to minimize the speckle intensity in a half dark hole going from~$-12.5\,\lambda/D$ to~$12.5\,\lambda/D$ in one direction and $2\,\lambda/D$ to~$12.5\,\lambda/D$ in the other direction. The coronagraphic image after five iterations is shown in~Fig.~\ref{fig:lab_images}~(d). The associated contrast curves calculated only inside the half dark hole are plotted in~Fig.~\ref{fig:lab_contrast}~(b). The~SCC correction in the half dark hole is better than in the full dark hole case (black curves in~Fig~\ref{fig:lab_contrast}~(a)). This is expected because using a single~DM in pupil plane (DM3) for the correction, both amplitude and phase aberrations can be corrected in a half dark hole whereas only phase aberrations can be controlled in a full dark hole. These encouraging results demonstrate that the SCC is capable of actively correcting static aberrations in long exposures reaching the residual limit set by the length of the exposure time and the level of AO residuals.

In Fig.~\ref{fig:lab_images}~(c) and~(d), we note that the remaining speckles are spatially modulated by the SCC fringes. We did not use this information in the current paper. However, it can be used to improve the contrast of these images using the post-processing coherence differential imaging mode of the~SCC \citep{sc}.

\section{Conclusion}
\label{s:con}

It is a well known problem that the long-exposure coronagraphic science images usually have an AO halo that adds photon noise to exoplanet detection and speckles that mimic exoplanet images. Static speckles with a temporal evolution longer than a typical angular differential imaging observing sequence can be subtracted in post-processing. However, quasi-static speckles that evolve slowly from one image to the other during the sequence of observations cannot be accurately calibrated post-observation. This paper focused on actively correcting the static and quasi-static speckles in long-exposure science images obtained under ground-based conditions. Given any focal plane wavefront sensor (FPWFS), we first established an expression of the fundamental accuracy on the measurement of static aberrations in a long-exposure in section~\ref{s:wave}. We then installed an atmospheric residual wheel on the THD2 bench mimicking post-AO residuals seen by the SPHERE/VLT instrument under good observing conditions. We then determined the level of static aberrations that can be measured from a finite long-exposure on THD2 in section~\ref{s:resultNum}. By using the self-coherent camera (SCC) as a FPWFS, the electric field was minimized down to the fundamental level set by the post-AO averaged turbulence.

It is shown in the laboratory that the SCC actively suppresses both phase and amplitude static aberrations when applied on AO long exposures. Only a few iterations are required to correct for the aberrations down to the fundamental limit, which means that quasi-static aberrations evolving on the scale of a few exposures could also be addressed with the SCC. A full dark hole with 1$\sigma$ contrast between $3\,\times\,10^{-6}\,$ and $\,8\,\times\,10^{-7}$ covering the region $5-12\,\lambda/D$ is reached in five iterations with images acquired at 18~seconds per iteration. In a half dark hole, a contrast of~$6\,\times\,10^{-7}$ is achieved in the range $12-15\,\lambda/D$. It is demonstrated that the SCC can actively compensate static and quasi-static aberrations present above the averaged turbulence. 

The results presented in this paper are encouraging and provide an opportunity to the current and future HCI instruments to adapt SCC as a FPWFS to actively suppress quasi-static speckles. We also aim to compare the SCC with other FPWFS techniques including the electric field conjugation on the THD2 bench. This study will characterize the performance of different speckle suppression techniques under the same ground-based conditions, thus setting a limit on the highest raw contrast obtainable from the ground. 

\begin{acknowledgements}

This project has received funding from the European Union’s Horizon 2020 research and innovation programme under the Marie Sk\l{}odowska-Curie grant agreement No 798909. This work has also received support of IRIS Origines et Conditions d’Apparition de la Vie of PSL Universit\'e under the program «Investissements d’Avenir » launched by the French Government and implemented by ANR with the reference ANR-10-IDEX-0001-02 PSL.  
\end{acknowledgements}

\bibliographystyle{aa} 
\bibliography{bib_GS}

\section*{Appendix: Variance of the averaged electric field in long exposure}
Here we describe in detail how the equations presented in section~\ref{s:wave} are derived. At a position~$\vec{\boldsymbol{\xi}}$ and at a time~$t$, the instantaneous electric field $\Psi(\vec{\boldsymbol{\xi}},\,t)$ at the entrance pupil in the presence of static aberrations~$\phi_0$ and evolving aberrations~$\phi_1(t)$ with a lifetime~$t_1$ can be represented as
\begin{equation*}
\Psi(\vec{\boldsymbol{\xi}},\,t) = P_0(\vec{\boldsymbol{\xi}})\,e^{\displaystyle i\,\left[ \phi_0(\vec{\boldsymbol{\xi}}) +\phi_1(\vec{\boldsymbol{\xi}},\,t)\right]} .
\label{eq:A1} \tag{A1}
\end{equation*}
The Eqs.~\ref{eq:psi_entrance} and~\ref{eq:A1} are the same. The terms~$\phi_0$ and~$\phi_1(t)$ can be described by Gaussian distributions with zero mean. The perfect coronagraph removes the coherent light \citep{cavarroc06} and the field~$\Psi_S(\vec{\boldsymbol{\xi}},\,t)$ in a pupil after the coronagraph can be written as
\begin{equation*}
\Psi_S(\vec{\boldsymbol{\xi}},\,t) = \Psi(\vec{\boldsymbol{\xi}},\,t) - P_0(\vec{\boldsymbol{\xi}})\,\iint \Psi(\vec{\boldsymbol{\xi}},\,t)\,\mathrm{d}\vec{\boldsymbol{\xi}}.
\label{eq:A2} \tag{A2}
\end{equation*}
Because there is a large number of different values of $\phi_0+\phi_1$ inside the pupil, the integral equals to the mathematical expectation. In this case, the expectation of~$e^{i\,\phi}$ is~$e^{-\sigma^2/2}$ with~$\sigma^2$ the statistical spatial variance of~$\phi$. Since~$\phi_0$ and~$\phi_1$ are statistically independent, we can write
\begin{equation*}
\Psi_S(\vec{\boldsymbol{\xi}},\,t) = P_0(\vec{\boldsymbol{\xi}})\,\left(e^{\displaystyle i\,\left[ \phi_0(\vec{\boldsymbol{\xi}}) +\phi_1(\vec{\boldsymbol{\xi}},\,t)\right]}-e^{\displaystyle-(\sigma_0^2+\sigma_1^2)/2}\right),
\label{eq:A3} \tag{A3}
\end{equation*}
where~$\sigma_i^2$ is the spatial variance of~$\phi_i$. Here, we assume that~$\sigma_1^2$ is constant over time. We note that~Eq.~\ref{eq:psi_s_stat} is a peculiar case of~\ref{eq:A3} (equivalent of~Eq.~\ref{eq:psi_s}) if~$\phi_1$ is null.

For an infinite exposure, the field~$\Psi_S$ averages over time and the resulting field is
\begin{equation*}
E\left[\Psi_S(\vec{\boldsymbol{\xi}},\,t)\right] = P_0(\vec{\boldsymbol{\xi}})\,\left(E\left[e^{ i\,\left[ \phi_0(\vec{\boldsymbol{\xi}}) +\phi_1(\vec{\boldsymbol{\xi}},\,t)\right]}\right]-e^{-(\sigma_0^2+\sigma_1^2)/2}\right),
\end{equation*}
where~$E[a]$ is the mathematical expectation of~$a$ over time. We then obtain
\begin{equation*}
E\left[\Psi_S(\vec{\boldsymbol{\xi}},\,t)\right] = P_0(\vec{\boldsymbol{\xi}})\,\left(e^{\displaystyle  i\,\phi_0(\vec{\boldsymbol{\xi}}) }\,e^{\displaystyle -\Sigma_1^2/2}-e^{\displaystyle -(\sigma_0^2+\sigma_1^2)/2}\right),
\end{equation*}
with~$\Sigma_1^2$ the time variance of~$\phi_1(\xi,\,t)$ that is assumed to be the same at each position~$\vec{\boldsymbol{\xi}}$ in the pupil. If~$\phi_1$ is ergodic, then~$\Sigma_1$ is equal to $\sigma_1$ and we obtain
\begin{equation*}
E\left[\Psi_S(\vec{\boldsymbol{\xi}},\,t)\right] = P_0(\vec{\boldsymbol{\xi}})\,\left(e^{\displaystyle i\,\phi_0(\vec{\boldsymbol{\xi}})}-e^{\displaystyle -\sigma_0^2/2}\right)\,e^{\displaystyle -\sigma_1^2/2}.
\label{eq:A4} \tag{A4}
\end{equation*}

\ref{eq:A4} is equivalent to Eq.~\ref{eq:psi_s_inf}. For a finite exposure, the resulting field~$<\Psi_S(\vec{\boldsymbol{\xi}},\,t)>_{\tilde{N}}$ is the average over~$\tilde{N}$ independent~$\phi_1(t)$ and can be written as
\begin{equation*}
    <\Psi_S(\vec{\boldsymbol{\xi}},\,t)>_{\tilde{N}} = \frac{1}{\tilde{N}}\,\sum_{p=1}^{p=\tilde{N}}\Psi_S(\vec{\boldsymbol{\xi}},\,p\,t_1).
\end{equation*}
We note here that the lifetime~$t_1$ of~$\phi_1$ can be different for each spatial frequency. For example, considering a single layer of frozen turbulent aberration moving at a speed~$v$ in front of the telescope of diameter~$D$, the lifetime~$t_{1,f}$ of the aberration with a spatial frequency~$f$ is proportional to $D/(f\,v)$. As a result, the expression of~$<\Psi_S(\vec{\boldsymbol{\xi}},\,t)>_{\tilde{N}}$ should be written in the Fourier space for a complete description. The resulting equations would be more complicated and will be presented in a future paper. This paper focuses on residual AO aberrations that are dominated by aberrations at low spatial frequencies (~$\lesssim2$ cycles per pupil diameter) and at the cut-off of the AO system (20 cycles per pupil diameter). We can then use the assumption that leads to the derived equation of~$<\Psi_S(\vec{\boldsymbol{\xi}},\,t)>_{\tilde{N}}$ for these two spatial frequencies neglecting the other frequencies.

The function$<\Psi_S(\vec{\boldsymbol{\xi}},\,t)>_{\tilde{N}}$ tends towards~$E\left[\Psi_S(\vec{\boldsymbol{\xi}},\,t)\right]$ if~$\tilde{N}$ increases. For a given~$\tilde{N}$, the variance of~$<\Psi_S(\vec{\boldsymbol{\xi}},\,t)>_{\tilde{N}}$ over time gives how much the finite average deviates from mathematical expectation. As all~$\Psi_S(\vec{\boldsymbol{\xi}},\,p\,t_1)$ are independent, we can write
\begin{equation*}
    \mathrm{Var}\left[<\Psi_S(\vec{\boldsymbol{\xi}},\,t)>_{\tilde{N}}\right] = \frac{1}{\tilde{N}}\,\mathrm{Var}\left[\Psi_S(\vec{\boldsymbol{\xi}},\,t)\right],
\end{equation*}
where~$\mathrm{Var}$ is the variance over time. We then use
\begin{equation}
    \mathrm{Var}\left[\Psi_S(\vec{\boldsymbol{\xi}},\,t)\right] = E\left[\left|\Psi_S(\vec{\boldsymbol{\xi}},\,t)\right|^2\right] - \left|E\left[\Psi_S(\vec{\boldsymbol{\xi}},\,t)\right]\right|^2.
\label{eq:A5} \tag{A5}
\end{equation}
The last term in the above equation can be written as
\small
\begin{equation}
    \left|E\left[\Psi_S(\vec{\boldsymbol{\xi}},\,t)\right]\right|^2 = P_0(\vec{\boldsymbol{\xi}})\,e^{-\sigma_1^2}\,\left[1+e^{-\sigma_0^2}-2\,e^{-\sigma_0^2/2}\cos{\left(\phi_0(\vec{\boldsymbol{\xi}})\right)}\right].
\label{eq:A6} \tag{A6}
\end{equation}
\normalsize
For~$E\left[\left|\Psi_S(\vec{\boldsymbol{\xi}},\,t)\right|^2\right]$, we first write~$\left|\Psi_S(\vec{\boldsymbol{\xi}},\,t)\right|^2$ as
\begin{equation*}
    \left|\Psi_S(\vec{\boldsymbol{\xi}},\,t)\right|^2 = P_0(\vec{\boldsymbol{\xi}})\,\left[1+e^{-\sigma^2}-2\,e^{-\sigma^2/2}\,\cos{\left(\phi_0(\vec{\boldsymbol{\xi}})+\phi_1(\vec{\boldsymbol{\xi}},\,t)\right)}\right],
\end{equation*}
with~$\sigma^2=\sigma_0^2+\sigma_1^2$. Then, we write the cosine as
\begin{equation*}
    \cos{\left(\phi_0+\phi_1\right)} = \cos{\phi_0}\, \cos{\phi_1} - \sin{\phi_0}\,\sin{\phi_1},
\end{equation*}
and as~$E[\phi_1]$ is null, the mathematical expectation over time gives
\begin{eqnarray*}
    E\left[\cos{\left(\phi_0(\vec{\boldsymbol{\xi}})+\phi_1(\vec{\boldsymbol{\xi}},\,t)\right)}\right]  &=& \cos{\left(\phi_0(\vec{\boldsymbol{\xi}})\right)}\, E\left[\cos{\left(\phi_1(\vec{\boldsymbol{\xi}},\,t)\right)}\right]\\
   E\left[\cos{\left(\phi_0(\vec{\boldsymbol{\xi}})+\phi_1(\vec{\boldsymbol{\xi}},\,t)\right)}\right]  &=&
   \cos{\left(\phi_0(\vec{\boldsymbol{\xi}})\right)}\,e^{-\sigma_1^2/2}.      
\end{eqnarray*}
We can then write
\small
\begin{equation}
  E\left[\left|\Psi_S(\vec{\boldsymbol{\xi}},\,t)\right|^2\right] = P_0(\vec{\boldsymbol{\xi}})\,\left[1+e^{-\left(\sigma_0^2+\sigma_1^2\right)}-2\,e^{-\left(\sigma_0^2/2+\sigma_1^2\right)}\, \cos{\left(\phi_0(\vec{\boldsymbol{\xi}})\right)}\right].
\label{eq:A7} \tag{A7}
\end{equation}
\normalsize
We derive the variance of~$\Psi_S(\vec{\boldsymbol{\xi}},\,t)$ from Eqs.~\ref{eq:A5}, \ref{eq:A6}, and~\ref{eq:A7}:
\begin{equation*}
    \mathrm{Var}\left[\Psi_S(\vec{\boldsymbol{\xi}},\,t)\right] = P_0(\vec{\boldsymbol{\xi}})\,\left(1-e^{-\sigma_1^2}\right).
\end{equation*}
Finally, the variance of~$<\Psi_S(\vec{\boldsymbol{\xi}},\,t)>_{\tilde{N}}$ is
\begin{equation*}
       \mathrm{Var}\left[<\Psi_S(\vec{\boldsymbol{\xi}},\,t)>_{\tilde{N}}\right] = P_0(\vec{\boldsymbol{\xi}})\,\frac{1-e^{-\sigma_1^2}}{\tilde{N}}.
       \label{eq:A8} \tag{A8}
\end{equation*}
\end{document}